\documentclass[amsmath,amsfonts,aps,pre,showpacs,twocolumn,superscriptaddress]{revtex4-1}

\usepackage[T1]{fontenc}
\usepackage[latin9]{inputenc}
\usepackage{graphicx}
\usepackage{color}
\usepackage{mathtools}
\usepackage{amsmath}
\usepackage{algorithm}
\usepackage[noend]{algpseudocode}
\usepackage{subcaption}

\makeatletter
\def\BState{\State\hskip-\ALG@thistlm}
\makeatother

\begin{document}

\title{Explosive Percolation on Directed Networks Due to Monotonic Flow of Activity}
\author{Alex Waagen}

\affiliation{HRL Laboratories, LLC., 3011 Malibu Canyon Rd, Malibu, CA 90265}
\author{Raissa M. D'Souza}
\affiliation{Department of Computer Science, University of California, Davis, CA 95616, USA}
\author{Tsai-Ching Lu}
\affiliation{HRL Laboratories, LLC., 3011 Malibu Canyon Rd, Malibu, CA 90265}

\begin{abstract}
An important class of real-world networks have directed edges, and in addition, some rank ordering on the nodes, for instance the ``popularity'' of users in online social networks. Yet, nearly all research related to explosive percolation has been restricted to undirected networks. 
Furthermore, information on such rank ordered networks typically flows from higher ranked to lower ranked individuals, such as follower relations, replies and retweets on Twitter. 
Here we introduce a simple percolation process on an ordered, directed network where edges are added monotonically with respect to the rank ordering.
We show with a numerical approach that the emergence of a dominant strongly connected component appears to be discontinuous. Large scale connectivity occurs at very high density compared with most percolation processes, and this holds not just for the strongly connected component structure but for the weakly connected component structure as well.  We present analysis with branching processes which explains this unusual behavior and gives basic intuition for the underlying mechanisms.  We also show that before the emergence of a dominant strongly connected component, multiple giant strongly connected components may exist simultaneously.  By adding a competitive percolation rule with a small bias to link uses of similar rank, we show this leads to formation of two distinct components, one of high ranked users, and one of low ranked users, with little flow between the two components. 

\end{abstract}

\maketitle

\section{Overview}
\label{sec:intro}

The percolation transition, corresponding to the emergence of large scale connectivity in networks, is of strong theoretical and practical interest~\cite{stauffer1985introduction,sahimi1994,Newman:2010:NI:1809753}. A recent focus has been on understanding mechanisms that lead to abrupt or ``explosive" percolation transitions and the consequences of such transitions~\cite{Achlioptas09, NPhys2015, boccaletti2016}.  Yet, limited work has explored abrupt percolation transitions on networks with directed edges~\cite{Squires13}, although there are important classes of real-world networks with directed edges, and moreover a rank ordering on the nodes. This is seen in online social networks such as Twitter where edges are directed (i.e., follower/followee edges) and the popularity of a user (i.e., number of followers) provides a natural ordering. Furthermore, the pattern of information flow is predominately from higher to lower ranked nodes, where less popular users tend to follow and share content from more popular users and not the reverse. That is, activity such as replies and retweets tend to flow in one direction with respect to popularity of the users. See Fig. 1 for a sample Twitter stream, which shows the characteristic pattern with only a small number of red edges represent links from more popular to less popular users. 

Inspired by such real-world networks, here we introduce two percolation models on a set of rank ordered nodes where edges are added monotonically with respect to the rank ordering, analogous to the monotonic flow of activity that tends to occur on online social networks such as Twitter. 

The first model, the ODER process, generalizes the directed {Erd\H{o}s-R\'{e}nyi} model to ordered graphs. This leads to the formation of two large components which then ``explosively" merge, showing that monotonic flow in a directed network is sufficient to yield an apparently discontinuous jump in the size of the largest strongly connected component. The second model, the C-ODER process, additionally incorporates competitive edge selection with a preference of connecting nodes of similar rank.   Again two large components emerge and eventually merge, but in a more ``explosive" manner. Yet, more surprising is that the small bias towards connecting users of similar rank leads to the two components separating users into two distinct classes. 
The two components have very little overlap in the rankings of the users, with one containing the lower ranked users and one containing the higher ranked users. 
Thus, a consequence of monotonic flow of information, with some bias towards grouping similar ranked nodes, leads to formation of two distinct groups of nodes which are divided by two classes with little flow of information between the classes.

The ODER process is simple enough that we can analyze it with branching processes, giving insight into the  fundamental underlying mechanisms.
We show that the branching processes die out very quickly due to the rank ordering imposed on the nodes in the network, and hence a high edge density is required to achieve even weak connectivity.  Furthermore, the monotonic nature of the ODER process with respect to the network ordering prevents the emergence of large strongly connected components. 
These two mechanisms which suppress large scale connectivity lay the groundwork for the sudden changes in the strongly connected component structure. 

The C-ODER adds a competitive rule for edge selection that enhances the abrupt nature of the ODER process. 
We will show that both the ODER and the C-ODER processes exhibit explosive growth in the size of the largest strongly connected component on a directed network.  
Moreover, we show compelling numerical evidence that the transition is discontinuous in the thermodynamic limit, as defined in Sec. \ref{cont}, and is the result of merging two giant strongly connected components. Finally, we demonstrate the unexpected feature of the C-ODER process, which is that there is very little overlap in the "intervals" spanned by the two giant components, which indicates that large components will contain users of similar rank. We define a component's interval by the set of all ranks between the lowest and highest ranked nodes contained in that component. In the ODER process, large components would contain randomly selected nodes, and hence the overlap between components' intervals would tend to be very large.

\subsection{Background}

Percolation is a pervasive mathematical concept describing the onset of large scale connectivity amongst nodes in a network with many applications in physics, chemistry, epidemiology, and complex networks~\cite{stauffer1985introduction, Newman:2010:NI:1809753}. We say that two nodes are in the same connected component if it is possible to reach one node from the other by successively following links. A network may consist of a single connected component, or be broken up into many distinct components.  A prototypical example of percolation, which we refer to as the  {Erd\H{o}s-R\'{e}nyi} process, begins with a collection of $n$ isolated nodes and sequentially adds undirected edges chosen uniformly at random from all possible edges~\cite{ER}. All components are initially of size one. As the number of edges increases and approaches $\frac{n}{2}$, a giant component (i.e., a component linear in system size $n$) emerges in a continuous, second order, phase transition.  

A directed version may be defined in which directed edges are added uniformly at random, again with a second order percolation transition occurring, but as the number of edges approaches $n$~\cite{bollobas1998random, newman2003structure}. In the case of directed random graphs, there are several different component structures to consider such as the strongly connected components, weakly connected components, in-components, and out-components, but the critical point is the same regardless of which component structure we consider. For definitions of these component structures see Sec. \ref{conndir}. 

In this manuscript we will study percolation in the context of a directed network, which adds both analytical and computational difficulty in tracking the overlapping component structures compared to the undirected case. Our particular focus is explosive percolation on real-world networks in which a natural ordering exists, and in which the formation of links tends to be monotonic with respect to that ordering.

Various modified versions of the  {Erd\H{o}s-R\'{e}nyi} process have been studied in order to gain a more sophisticated understanding of how networks form. Recently, many such modified processes have employed a competitive dynamic in which multiple candidate edges are selected at each discrete time step, but only one is actually be added to the network. The criterion used to select the winning edge typically considers the sizes of the components that would be joined by the edge. Such competitive percolation models first appeared in 2009, with the introduction of the \textquotedbl{}product rule\textquotedbl{}, where two edges are chosen at random each discrete time step, but only the edge that minimizes the product of the components to be merged is actually added to the graph~\cite{Achlioptas09}. Such a process appeared to lead to a discontinuous percolation transition, although it was later shown in 2011 that any rule with a fixed number of competitive edges ultimately leads to a continuous percolation transition in the thermodynamic limit~\cite{Riordan15072011}, yet the universality class of the transition is extremely unusual~\cite{PhysRevLett.105.255701,grassberger2011,lee2011}.  Much more is now understood about ``explosive" percolation transitions, including the difference between edge-competition and node-competition in addition to many variants that display discontinuous transitions in the thermodynamic limit. For a recent review discussing these issues see Ref.~\cite{NPhys2015}. Note that a discontinuous emergence of large scale connectivity is characterized by a change in the size of the largest component by $\theta(n)$ as a result of adding $o(n)$ edges.

In 2013, Squires et al.~\cite{Squires13} adapted the product rule to directed networks and showed that the weakly connected component along with the largest in-components and out-components exhibit sudden growth, though not quite as sudden as the transition exhibited by the product rule in undirected networks. They refer to this as ``weakly explosive'' percolation. However, the emergence of a giant strongly connected component is clearly continuous even for relatively small system sizes. Note that the strongly connected component structure can be crucial to the flow of activity on a network. It allows activity not just to flow outwards and dissipate throughout the network, but to return and be reinforced. Moreover, imposing an ordering on the nodes adds inherent meaning to the directed links beyond topological structure, and may result in a model which more closely resembles reality in some cases.

\subsection{Motivation}

In this paper we will define two processes. The first process is referred to as the ODER (ordered, directed Erd\H{o}s-R\'{e}nyi) process, and in this process directed edges (a,b) are added uniformly at random under the constraint that $a$ precedes $b$ with respect to the ordering. The edge $(b,a)$ will be added in the event that $(a,b)$ already exists in the graph. The C-ODER process is a modification of this process which utilizes edge competition, in that at each step the prospective edges between nodes a,b,c are considered and only one edge is selected to be added to the graph. 

Since the probability of choosing a given edge $(a,b)$ twice is small, it is clear that until a high density has been achieved, in the ODER and C-ODER processes edges will almost always be formed from a lower ranking node to a higher ranking node. See Sec. \ref{sec:analysis} for a proof that the expected number of reverse edges grows as the square of the edge density. This has an analogue in some real world directed networks such as Twitter, in which it is intuitive that individuals with less influence will usually follow user with more influence and not the reverse. 

\begin{figure*}
\begin{subfigure}{0.31\textwidth}
\includegraphics[scale=0.35]{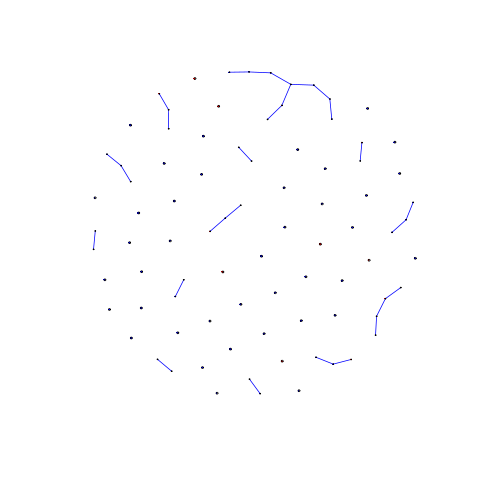}
\captionsetup{justification=raggedright,
singlelinecheck=false
}

\end{subfigure}
\begin{subfigure}{0.31\textwidth}
\includegraphics[scale=0.35]{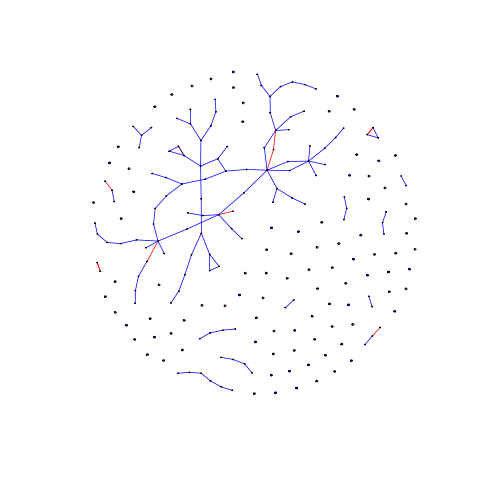}
\captionsetup{justification=raggedright,
singlelinecheck=false
}

\end{subfigure}
\begin{subfigure}{0.31\textwidth}
\includegraphics[scale=0.35]{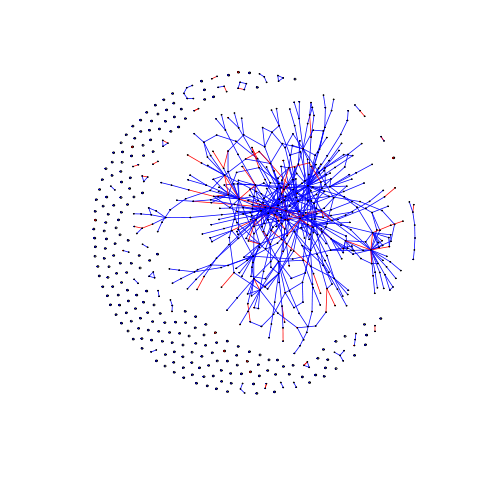}
\captionsetup{justification=raggedright,
singlelinecheck=false
}

\end{subfigure}
\captionsetup{justification=raggedright,
singlelinecheck=false
}
\caption{Temporal evolution of the retweet network on Twitter after the discovery of the Higgs Boson. Blue links indicate a lower rank node retweeting a higher ranked node, and red links indicate a higher ranked node retweeting a lower ranked node. Nodes of degree 0 and 1 are removed for the sake of readability, and any nodes that appear to be of degree 0 or 1 are actually connected to some number of degree 1 nodes that were removed. (Left) First 1000 retweets. (Middle) First 4000 retweets. (Right) First 10000 retweets. This dataset is publicly available as part of the Stanford Large Network Dataset Collection.\cite{snapnets}}
\label{fig:higgs}   
\end{figure*}

\begin{figure}

\includegraphics[scale=0.3]{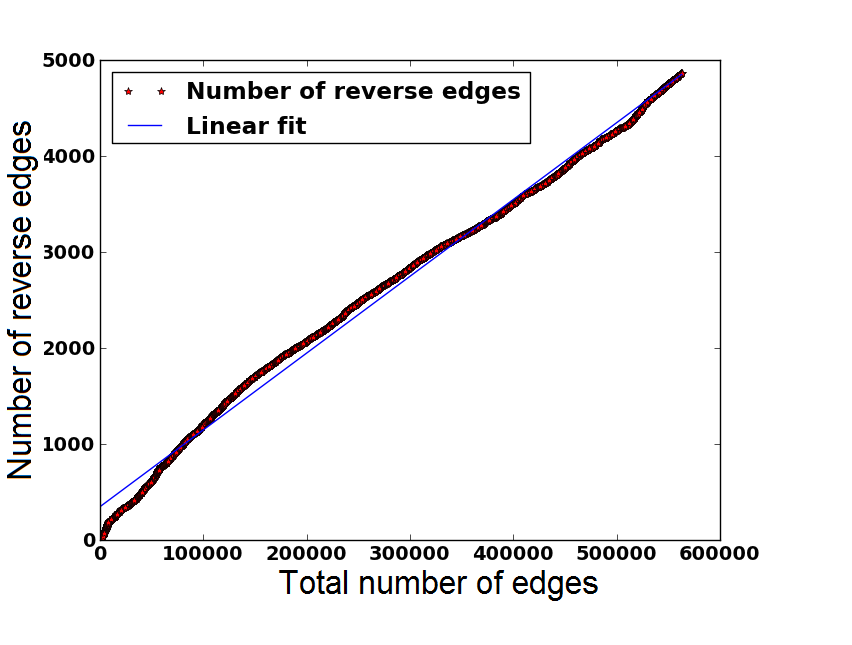}
\captionsetup{justification=raggedright,
singlelinecheck=false
}
\caption{The number of ``reverse edges'' compared to the total number of edges. That is, the number of times a higher ranked user retweets a lower ranked user compared to the total number of retweets after the discovery of the Higgs Boson. Note that the number of reverse edges is only about 1\% of the total number. The optimal linear fit is given by $y = 0.008x + 347$.}
\label{fig:reverse_edges}   
\end{figure}

Suppose that we rank users on Twitter according to, say, total number of followers, total number of retweets, Pagerank, or some other measure of influence. There are many ways to rank users on Twitter according to their influence, and it is a topic that has been addressed many time \cite{bakshy2011everyone, weng2010twitterrank, kwak2010twitter, leavitt2009influentials, cha2010measuring}. In Fig. \ref{fig:higgs} users are ranked by the number of total retweets and red edges denote a higher ranked user retweeting a lower ranked user. The number of red edges is about 1\% of the total edges, as seen in Fig. \ref{fig:reverse_edges}.  
Past studies find that a user of medium rank is often a popular internet personality and a user of high rank may be a politician or celebrity.  A low rank user is hence likely to follow medium and high ranked users, and a medium rank user is likely to follow a high ranked user, but the reverse is unlikely. For instance, Twitter organization accounts (news media, schools, entertainment media, etc.) often exist solely to broadcast information to their followers \cite{park2015network}.   Moreover, not all links may be meaningful over a short time frame due to the bursty dynamics of Twitter which result in many information links being quickly created and destroyed~\cite{myers2014bursty}. By analyzing user activity, it is possible to infer different network structures depending on which links we consider meaningful~\cite{de2010inferring, ma2014inferring, gilbert2015tie}. In a short time frame, it 
it is often the case that almost all user activity consists of interaction between similarly ranked users or with lower ranked users retweeting or responding to tweets from higher ranked users. This is evident in the Higgs Boson Twitter dataset, which includes only activity immediately following the discovery of the Higgs Boson particle.    

\subsection{Outline}

In Sec. \ref{sec:prelim} we define some basic concepts, in Sec. \ref{sec:growth} we define the ODER and C-ODER processes in detail, in Sec. \ref{sec:numerical} we show numerically that the C-ODER process leads to the formation of two giant strongly connected components which merge to form a dominant strongly connected component, and in Sec. \ref{sec:analysis} we analyze the ODER process to explain the high critical density required to achieve large scale strong connectivity.

\section{Preliminaries}
\label{sec:prelim}

\subsection{Continuity of phase transitions}
\label{cont}

We say that a phase transition is continuous if a positive change in edge density always results in a positive change in the size of the largest connected component relative to system size. To clarify this, we define the edge density $\delta = \frac{m}{n}$  (where $m$ is the number of directed edges and $n$ is the number of nodes), and the size of the largest component as $C_1(\delta)$. Hence the size of the largest component relative to system size is $\frac{C_1(\delta)}{n}$, and a phase transition is continuous if with high probability $\lim_{\epsilon \rightarrow 0} \lim_{n\rightarrow \infty} \frac{C_1(\delta + \epsilon)}{n} = \lim_{n\rightarrow \infty} \frac{C_1(\delta)}{n}$ for any density $\delta$. Conversely, we say that a percolation phase transition is discontinuous if it is not continuous.  


\subsection{Connectivity structure in directed networks}
\label{conndir}
Directed networks have a fundamentally different component structure when compared to undirected networks, so it is crucial 
to understand how percolation, and especially explosive percolation, may occur differently. On directed networks, the notion of connectivity is more complicated, because the situation arises in which node $x$ may reach node $y$ by following successive edges but the reverse is not true. There are, therefore, many different ways of defining connectivity on directed networks. For convenience, let us use the notation $x \sim y$ if it is possible to travel from node $x$ to node $y$ by following successive edges. We define the following:
\begin{enumerate}
\item The strongly connected component $SCC(x)$ containing $x$ is the node $x$ together with the set of all nodes $y$ which satisfy $x \sim y$ and $y \sim x$. 
\item The weakly connected component $WCC(x)$ containing $x$ is the node $x$ together with the set of all nodes $y$ which satisfy $x \sim y$ or $y \sim x$. 
\item The out-component $OUT(x)$ containing x is the node $x$ together with the set of nodes $y$ which satisfy $x \sim y$.
\item The in-component $IN(x)$ containing x is the node $x$ together with the set of nodes $y$ which satisfy $y \sim x$.
\end{enumerate}

\subsection{Ordered graphs}

An ordering may be defined on any countable set $S$ via a function $f: S \rightarrow \mathbb{N}$. If $i,j \in S$ we say that $i$ is lower or equal in the ordering if and only if $f(i) \leq f(j)$. We may represent this symbolically as $i \preceq j$.  
Without loss of generality, suppose that the ordering is a function of the form $f : N(G) \rightarrow \mathbb{N}$, so that given two nodes labeled as $i$ and $j$, $i \preceq j$ if and only if $f(i) \leq f(j)$. In this paper we place an arbitrary ordering on the nodes and also label the nodes arbitrarily. Hence without loss of generality we may label the nodes with the natural numbers and define an ordering as $f(i) = i$. That is, their place in the ordering is the same as their label, and hence $i \preceq j$ if and only if $i \leq j$. We refer to the node labeled $i$ as the node ranked $i$, and we say that $j$ is a higher ranked node than $i$ if $i \preceq j$.

\subsection{Intervals and overlap}
\label{interval}

With the addition of ordering to a network, it becomes meaningful to discuss the location of components in the ordering. Here we define the interval spanned by a component and the overlap between two components.

Let $C$ be a strongly connected component in network $G$. If $a$ is the rank of the lowest ranked node in $C$ and $b$ is the rank of the highest ranked node in $C$, then we say that the real interval $[a,b]$ is the interval spanned by $C$. If $[a,b]$ is the interval spanned by component $C$ and $[c,d]$ is the interval spanned by component $D$, then the overlap between components $C$ and $D$ is the length of the intersection $[a,b] \cap [c,d]$. Note that in the standard ER process with an arbitrary ordering on the nodes, any large component will with high probability span almost the entire interval, and hence, that a related process (i.e. C-ODER) can result in a small overlap between two giant components, is notable.

\subsection{Algorithms to track component size}

With the more complicated connectivity structure of directed graphs comes more algorithmic complexity in keeping track of component sizes. In undirected networks it is possible to quickly update the component structure with the addition of each additional edge, so that we can track the component structure throughout the entire process with $O(1)$ operations for each edge addition, for a total of $O(n)$ operations. This is done by denoting a ``root'' node for each component, and updating the root node each time two components are merged. This process is known as the Newman-Ziff algorithm \cite{newman2001fast}. However, in directed networks this method is not effective. In \cite{Squires13} the component structure is tracked throughout the process with an original method which requires approximately $O(n^{1.5})$ operations, but here we will use a method which only requires $O(E\log E)$ operations, where $E$ is the total number of edges in the system after the process has been halted. 
For our method we only track the component structure near critical points where a large jump is observed in the size of the largest strongly connected component, and we use a standard method~\cite{tarjan1972depth} to calculate the strongly connected components at these points. More details will be given in Sec. \ref{sec:numerical}.

\section{Growth Processes on Ordered, Directed Networks}
\label{sec:growth}

In this section we present two growth processes which form ordered, directed networks. The ODER process is a natural extension of the Erdos Renyi process to ordered, directed networks, and the C-ODER process modifies the ODER process with a competitive rule. 

\subsection{Ordered, Directed {Erd\H{o}s-R\'{e}nyi}}

\begin{figure}
\begin{subfigure}{0.23\textwidth}
\includegraphics[scale=0.5]{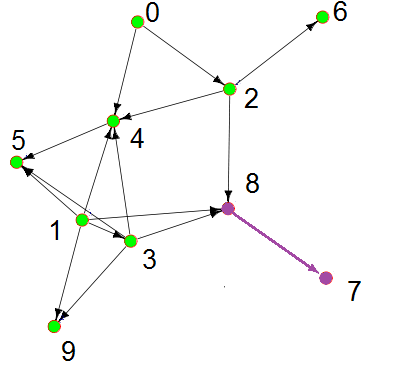}
\captionsetup{justification=raggedright,
singlelinecheck=false
}
\end{subfigure}
\begin{subfigure}{0.23\textwidth}
\includegraphics[scale=0.5]{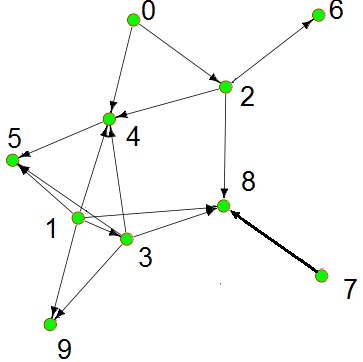}
\captionsetup{justification=raggedright,
singlelinecheck=false
}

\end{subfigure}
\captionsetup{justification=raggedright,
singlelinecheck=false
}
\caption{In the ODER process directed edges $(a,b)$ are selected uniformly at random.  If $a<b$ in ranking, then the edge $(a,b)$ is added to the network, and otherwise the edge $(b,a)$ is added. (Left) The purple-colored edge $(8,7)$ is chosen at random. (Right) Since $8 > 7$ the edge $(7,8)$ is added in the right image.}
\label{fig:fig1}   
\end{figure}

We begin with a set of $n$ isolated nodes on which we have placed an arbitrary ordering from $1$ to $n$, and at each time step a single directed edge will be added between two nodes selected uniformly at random. Moreover, the head of the directed edge will always be the node which is higher in the ordering unless the edge already exists in the graph. In that case the head of the edge will be the node which is lower in the ordering. See Fig. \ref{fig:fig1}.

This process is repeated until $m$ edges have been added to the graph. For instance, if $n = 10$ we initialize a set of $10$ isolated nodes which are labeled from $0$ to $9$.  For convenience, we interchangably refer to the ranking and labeling of nodes. That is, the label of a node is also its rank in the ordering.  In the left image of Fig. \ref{fig:fig1}, 13 edges have been added to the graph. In order to determine the 14th edge we choose two random nodes, node $8$ and node $7$. The prospective directed edge $(8,7)$ is colored red. However, since $8>7$, instead the edge $(7,8)$ is added to the network. This is shown in the right image of Fig. \ref{fig:fig1}. If we think of these nodes as representing users on Twitter and the each individual edge represents a one-way social connection, then the addition of $(7,8)$ corresponds to an event in which user $7$ retweets user $8$. If later in the process either the edge $(7,8)$ or $(8,7)$ is chosen, then the reverse edge $(8,7)$ will be added to the network. 

\begin{figure}
\includegraphics[scale=0.75]{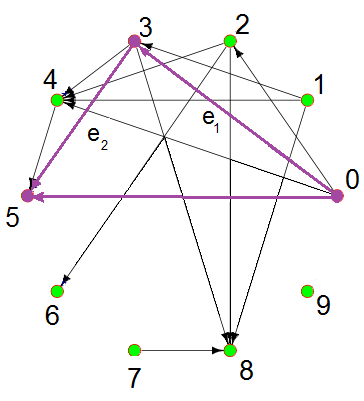}
\captionsetup{justification=raggedright,
singlelinecheck=false
}
\caption{In the C-ODER process three random nodes $a<b<c$ are selected, and either the edge $e_1 \coloneqq (a,b)$ or the edge $e_2 \coloneqq (b,c)$ is added to the network. If $b-a < c-b$ we add $e_1$, and otherwise we add $e_2$. In the image above the edge $e_2$ will be selected. Regardless of the particular values of a,b, and c, since $a<b<c$ it follows that $c - a > b - a$ and $c - a > c - b$, so the edge $(a,c)$ will never be selected. In the figure a=0, b=3, and c =5. Since 5 - 3 < 3 - 0. the edge $e_2 \coloneqq (3,5)$ is added iinstead of the edge $e_1 \coloneqq (0,3)$.}  
\label{fig:fig3}
\end{figure}

\subsection{Competitive Ordered, Directed {Erd\H{o}s-R\'{e}nyi}}
\label{sec:CODER}

In order to facilitate a more sudden emergence of a dominant SCC, we alter the ODER process with a competitive rule as follows. At each timestep three nodes, $a<b<c$, will be chosen uniformly at random. Among these nodes are three candidate edges $(a,b)$, $(a,c)$, and $(b,c)$, only one of which will be added to the graph. The edge to be added will be the one which minimizes the distance between the head and tail nodes. Note that the edge (a,c) will never be added, because $c - a > b - a$ and $c - a > c - b$.  For instance, if $a = 0$, $b = 3$, and $c = 5$, then the edge $(b,c)$ will be selected because $c - b = 5 - 3 =2$ is less than $b - a = 3 - 0 = 3$ and $5 - 0 = 5$. This is illustrated in Fig. \ref{fig:fig3} in which $e_1 \coloneqq (0,3)$ and $e_2 \coloneqq (3,5)$ are the edges $(a,b)$ and $(b,c)$. 

Note that by choosing the edge which minimizes the difference in rank between the two nodes we not only encourage links between users of similar rank, but discourage lower ranked users from following higher ranked users. In fact, it is impossible for the difference in rank to be more than $\frac{n}{3}$, where $n$ is the number of nodes in the system. If we take inspiration from Twitter this may seem somewhat problematic, because many low-ranked users follow users of much higher rank. A more realistic process would incorporate both the tendency of users to interact with those of similar rank and also the tendency for lower ranked users to follow users of much higher rank with little need for interaction. However, our goal is not to exhibit a single process which serves as a model for user activity on Twitter, but to isolate and study mechanisms which lead to interesting behavior, particularly a discontinuous jump in the size of the largest strongly connected component. 

The C-ODER process is illustrated in Fig. \ref{fig:fig3}. The three chosen nodes are $0$, $3$, and $5$, so that the three possible edges are $(0,3)$, $(0,5)$, and $(3,5)$. These prospective edges are denoted by thick, purple arrows, while edges which have been previously added are denoted by black arrows. One of the three purple edges will be added to the graph, and the other two will be discarded. We keep the edge in which the distance in the ordering between the head and the tail of the edge is smaller. Since the difference between $3$ and $5$ is less than the difference between $1$ and $3$ or $1$ and $5$, we add the edge from $3$ to $5$.

\section{Numerical Analysis of the C-ODER Process}
\label{sec:numerical}

\begin{figure}
\includegraphics[scale=0.35]{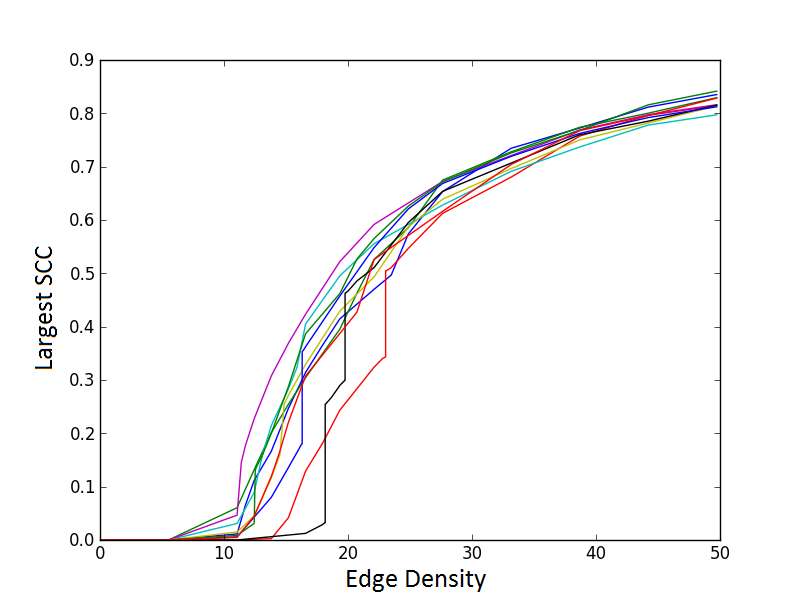}
\captionsetup{justification=raggedright,
singlelinecheck=false
}
\caption{The ODER process over $10$ different runs of $1,000,000$ node networks. Behavior varies greatly in different trials, and so it is not clear whether explosive percolation occurs. In particular, the analysis in Sec. \ref{sec:analysis} suggests that the initial emergence of a giant SCC may be discontinuous.}
\label{fig:fig2million}   
\end{figure}

\begin{figure}
\includegraphics[scale=0.35]{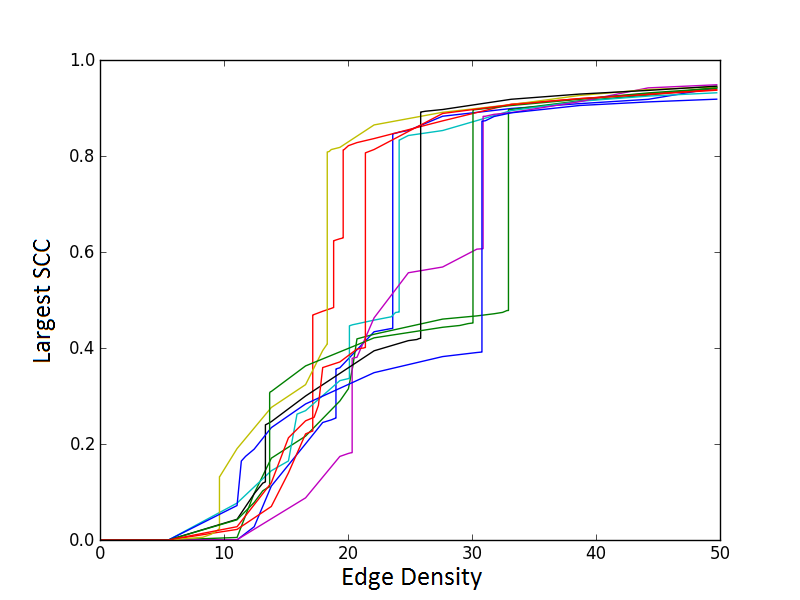}
\captionsetup{justification=raggedright,
singlelinecheck=false
}
\caption{Illustration of explosive percolation (discontinuous jump) with the C-ODER process over $10$ different runs of $1,000,000$ node networks.} 
\label{fig:fig4million}   
\end{figure}

In this section we analyze the C-ODER process and show that not only is the emergence of a dominant SCC discontinuous, but it occurs as a result of merging two giant SCCs with very small overlap as defined in Sec. \ref{interval}. First, see Fig. \ref{fig:fig2million}, which shows that in the ODER process it is often the case that no discontinuous jumps in the size of the largest SCC occur. The behavior in the ten different trials varies greatly, and it is not clear whether or not explosive percolation occurs. However, in Fig. \ref{fig:fig4million}, which shows a similar plot with the C-ODER process, every trial exhibits a clear discontinuous jump resulting from the addition of a single edge.

Fig. \ref{fig:fig4million} shows the fractional size of the largest strongly connected component as density is increased in $10$ different trials with $n = 1,000,000$. It appears to be very likely that neither the points where we see large jumps nor the size of these jumps will converge to a single value in the thermodynamic limit. This stochastic nature of the jumps is seen in several known models of explosive percolation~\cite{NPhys2015}.
In order to quantify the size of the jumps, it is therefore necessary to take the average of the largest jumps over many different trials.

\begin{figure}
\begin{subfigure}{0.31\textwidth}
\includegraphics[scale=0.45]{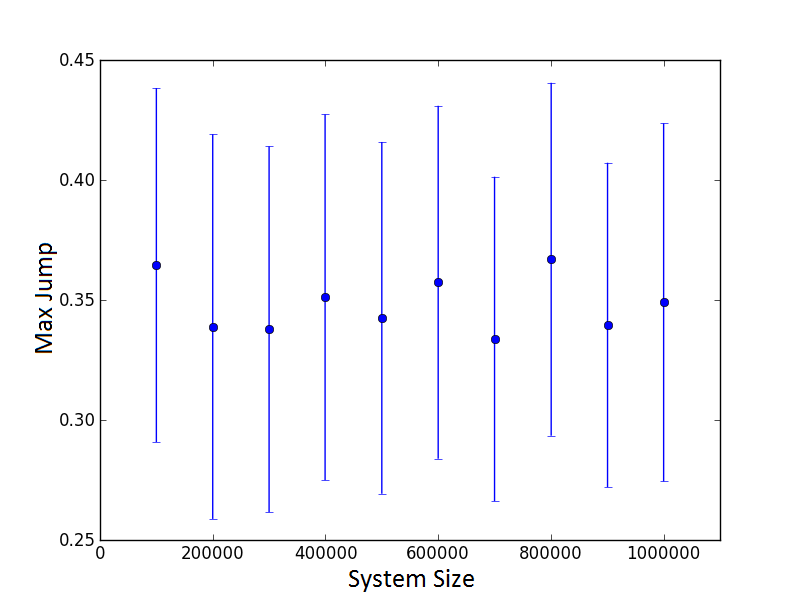}
\captionsetup{justification=raggedright,
singlelinecheck=false
}

\end{subfigure}
\begin{subfigure}{0.31\textwidth}
\includegraphics[scale=0.45]{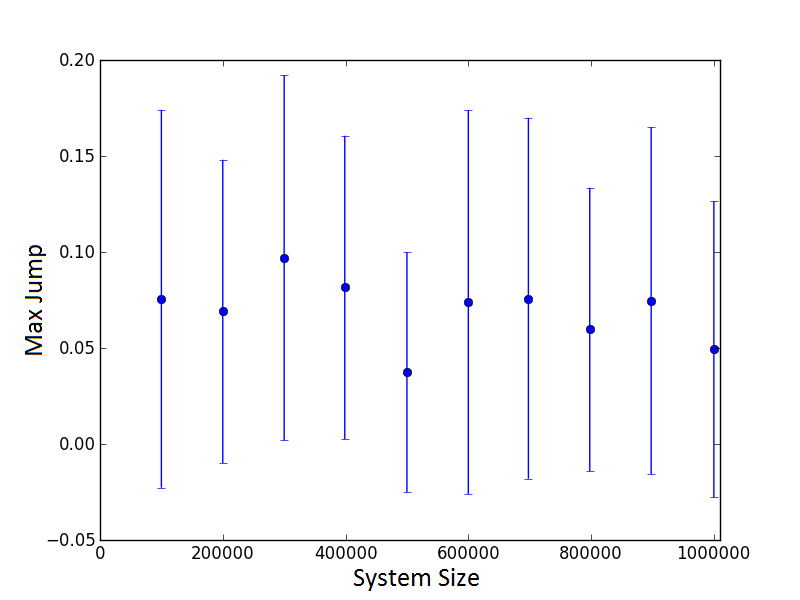}
\captionsetup{justification=raggedright,
singlelinecheck=false
}
\end{subfigure}   
\caption{The maximum jump in the size of the largest strongly connected component resulting from the addition of a single edge averaged over $40$ runs. As system sizes increases, the size of the jump remains constant. (Top) In the C-ODER process, there is a large jump about $\frac{1}{3}$ of the system size in magnitude. (Bottom) In the ODER process, there is a smaller jump about $5\%$ of the system size in magnitude.}
\label{fig:fig6million}
\end{figure}

\begin{figure}
\includegraphics[scale=0.45]{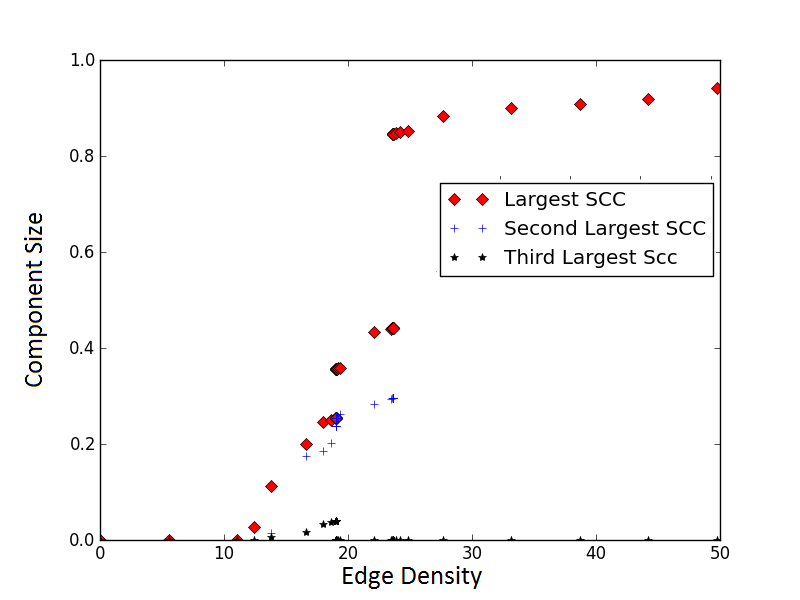}
\captionsetup{justification=raggedright,
singlelinecheck=false
}
\caption{The three largest strongly connected components from a single run. Here we see that two giant strongly connected components can exist simultaneously until they are merged into a single larger component.}
\label{fig:fig7million}   
\end{figure}

\subsection{Is the jump discontinuous?}
\label{jumpdis}

Although the jumps we observe in Fig. \ref{fig:fig4million} over $10$ independent trials appear to be discontinuous, it may be the case that in very large systems the size of the jump eventually becomes sublinear. A possible numerical approach is to increase the system size and see if the jump appears to be increasing linearly. However, this approach is not reliable for detecting discontinuity. In \cite{Achlioptas09} it was shown that the size of the jump appeared to be linear even when a number of edges on the order of $n^{\frac{2}{3}}$ were added near the critical point. However, it was later proved rigorously~\cite{Riordan15072011} that the transition is continuous in the thermodynamic limit. For this reason, we require that the jump results from the addition of a single edge~\cite{nagler2011}. However, this is not a proof of discontinuity, since it may simply be that the critical exponent is too small for the continuous nature to be detected on systems of this size. For this reason we also analyze the ODER process in Sec. \ref{sec:analysis} in order to gain intuitive understanding of how a discontinuous jump may occur via the emergence of large scale weak connectivity before any significant strong connectivity structure exists.

Define $\Delta(t) \coloneqq C_1(\frac{t}{n}) - C_1(\frac{t-1}{n})$ where $C_1(\frac{t}{n})$ is the size of the largest component after the addition of $t$ edges, as defined in Sec. \ref{sec:prelim}. That is, $\Delta(t)$ is the size of the jump of $\frac{C_1}{n}$ resulting from the addition of the $t$th edge. Let $M \coloneqq \max_{t>0} \Delta(t)$. Then $M$ is the largest jump in the size of the largest component throughout the entire process. We wish to show that $M>cn$  for some $c>0$ as $n\rightarrow \infty$.
\section{Calculating the Largest Jump}
\label{jumpcalc}
The simplest way to track the size of the largest strongly connected component is to simply recalculate the component structure after each edge addition, using for example Tarjan's algorithm \cite{tarjan1972depth}. Tarjan's algorithm runs in time linear in the number of edges, so the largest SCC can be tracked over $O(n\log{n})$ edge additions in $O(nE\log{n})$ time. A trivial way to speed up this calculation of the component structure is to only recalculate the component structure every $k$ edge additions, ignoring the intermediary edge additions. This speeds up the algorithm by a factor of $k$, but obviously makes calculation of the largest jump inaccurate since many edge additions will occur in which we do not track the size of the largest SCC. If we take $k=\frac{n\log n}{2}$ then we will only calculate the component structure twice, resulting in an algorithm that runs in linear time.Obviously, this does not yield the size of the largest jump resulting from a single edge addition. However, it give us information on whether any large jumps occur ingive  $[0,\frac{n\log{n}}{2}]$ or $[\frac{n\log{n}}{2}+1,n]$. If the difference in the size of the largest strongly connected component at the beginning and end of these intervals is sufficiently small, say less than $0.01n$, then we may discard the interval. Otherwise, we keep the interval and proceed with a binary search. If a jump occurs in both intervals, then we must search both, and this process proceeds recursively. For more details, see the pseudocode in Algorithm 1. So long as the number of jumps of size $0.01n$ or greater is finite, the number of binary searches needed will be bounded by a constant, so that we can find the location of the largest jump of size $0.01n$ or greater in $O(E\log E)$ operations. Here, $E$ refers to the total number of edges in the system after the process is halted.

\begin{algorithm}[H]
\captionsetup{justification=raggedright,
singlelinecheck=false
}
\caption{Get Largest Jump}
\begin{algorithmic}[1]
\Procedure{GetLargestJump}{}
\State $\textit{SCC1(\textit{L},\textit{q})} \gets \text{largest SCC from edge list L[0:q]}$
\State $Edges \gets \textit{empty ordered list}$
\State $m \gets \textit{number of edges to add}$

\For {$i \textit{ ranging from } 1 \textit{ to } m$}
\State $(u,v) \gets \textit{ edge chosen uniformly at random}$
\If {$b > a$}
\If {$(a,b) \textit{ does not exist }$} 
\State $ \textit{ append} (a,b)  \textit{ to } Edges$
\Else
\State $ \textit{ append} (b,a)  \textit{ to } Edges$
\EndIf
\Else
\If {$(b,a)  \textit{ does not exist}$} 
\State $\textit{ append } (b,a) \textit{ to } Edges$
\Else
\State $\textit{ append } (a,b) \textit{ to } Edges$
\EndIf
\EndIf
\EndFor

\State $head \gets SCC1(Edges,1)$
\State $tail \gets SCC1(Edges,m)$
\State $mid \gets SCC1(Edges,m/2)$
\State $LargestJump \gets 0$

\If {$tail - head < n/100$} \Return LargestJump 
\EndIf

\If{$mid - head  > n/100$}
\State{$BinSearch(start, m/2)$}
\EndIf

\If{$tail - mid  > n/100$}
\State $BinSearch(m/2, m)$
\EndIf

\Return LargestJump

\EndProcedure
\end{algorithmic}
\end{algorithm}

\begin{algorithm}[H]
\captionsetup{justification=raggedright,
singlelinecheck=false
}
\caption{Binary Search}
\begin{algorithmic}[1]
\Procedure{BinSearch(start,end)}{}

\State $head \gets SCC1(Edges, start)$
\State $tail \gets SCC1(Edges,end)$
\State $mid \gets SCC1(Edges,(start + end)/2)$

\If {$end - start = 1$}
\If {$tail - head > LargestJump$}
\State $LargestJump \gets tail - head$	
\EndIf
\EndIf		
\If{$mid - head  > n/100$}
\State $Binsearch(start, mid)$
\EndIf
\If{$tail - mid  > n/100$}
\State $Binsearch((start + mid)/2, tail)$
\EndIf

\EndProcedure
\end{algorithmic}
\end{algorithm}

\subsection{Evidence of discontinuity}
\label{dis}

In order to produce the plot in Fig. \ref{fig:fig6million} we run $40$ different trials of the C-ODER process for each system size $n$, for $n=100,000$ to $1,000,000$, and plot the mean average size of $M$ for each system size over the $40$ independent trials. We also plot the standard deviation over these $40$ trials as error bars. As system size increases, the size of the largest jump is proportional to system size at approximately $\frac{n}{3}$ regardless of system size. For a process with critical exponent $\beta$ the size of the largest maximum jump grows as $n^{-\beta}$\cite{nagler11}. Since the plot in Fig. \ref{fig:fig6million} does not show any decrease, the implication is that $\beta = 0$, implying a discontinuous jump, or $\beta$ is sufficiently small that the decrease is not significant. Note that although the largest jump often does not coincide to the initial emergence of large scale connectivity, it may still be the case that the initial emergence of a giant strongly connected component is discontinuous. This separate question is addressed in Sec. \ref{sec:analysis} with respect to the ODER process, which is more amenable to analysis than the C-ODER process.

\subsection{Two co-existing giant strongly connected components}
\label{coexist}
Fig. \ref{fig:fig7million} shows that two giant strongly connected components can exist up until the point when a single strongly connected component dominates the network. We see that the discontinuous jump in the size of the dominant SCC coincides with the disappearance of the second largest SCC. Note that merging two strongly connected components may result in a component larger than the sum of all nodes contained in both components. If a node is in the in-component of one SCC and the out-component of another SCC, then merging those two SCCs will cause that node to be part of the newly merged SCC. That is, suppose $A$ and $B$ are strongly connected components. Then the SCC which results from merging those two components is $A \cup B \cup (OUT(A) \cap IN(B)) \cup (IN(A) \cap OUT(B))$, which can be significantly larger than $A \cup B$. This is demonstrated in Fig. \ref{fig:fig7million}, in which we see two giant SCCs merging to create a single, much larger SCC.  

\begin{figure}
\includegraphics[scale=0.45]{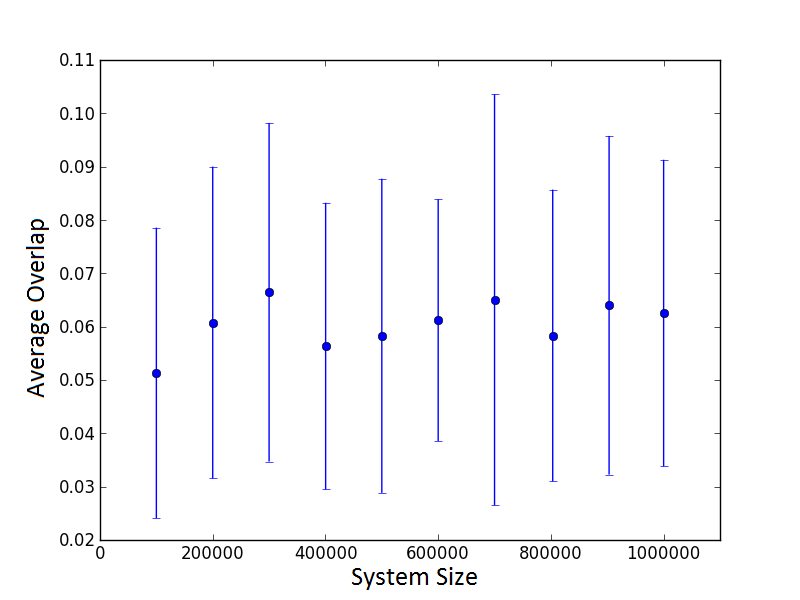}
\captionsetup{justification=raggedright,
singlelinecheck=false
}
\caption{The fractional overlap between the intervals of the two largest strongly connected components in the C-ODER process immediately prior to the largest jump in the size of the largest strongly connected component, averaged over $20$ independent trials. }
\label{fig:fig8million}   
\end{figure}

Moreover, the two largest SCCs tend to have very little "overlap" among their nodes. As defined in \ref{interval}, the interval spanned by $A$ is the interval $[a,b]$ where $a$ is the lowest ordered node in the set $A$ and $b$ is the highest ordered node in the set $A$. When we say there is very little overlap in the the first and second largest SCCs, $C_1$ and $C_2$, we mean that there is a small overlap in the intervals spanned by $C_1$ and $C_2$.   In Fig. \ref{fig:fig8million} we see that the overlap averaged over $20$ independent trials is around 5\% of system size. The average length of the interval spanned by $C_1$  is about $0.6n$, and the average length of the interval spanned by $C_2$ is about $0.4n$. Note that there is no reason for these components to be separated merely as a consequence of being distinct strongly connected components. Indeed, if we were to apply a random permutation to the current ordering, it is straightforward to see that with high probability the resulting overlap would approach $n$ in the thermodynamic limit.

It is not entirely clear why this extremely small overlap occurs, but it would be interesting to know if similar behavior appears in real-world directed networks which have a natural ordering. For example, similar properties may be observed in the user activity graph on Twitter over a short time frame.  That is, if some natural ranking is placed on the nodes and connectivity among the nodes is defined in some way, there may be some point where two or three strongly connected components exist, containing nodes with very little overlap amongst their intervals. To be concrete, the ranking could be the number of total retweets a user has had over their entire lifespan, and the edge $(a,b)$ could indicate user $a$ retweeting user $b$ within the time frame of interest. 

\section{Analysis of the ODER Process}
\label{sec:analysis}

In this section we analyze the simpler ODER process in order to determine why such a high edge density is required in order to achieve large scale connectivity. First we use branching process analysis to show that giant in-components and giant out-components do not exist until the edge density is of order $\log(n)$ so long as reverse edges are ignored, and then we will show that reverse edges have an insignificant effect on the component structure while in the subcritical phase.

\subsection{Number of reverse edges added}
\label{reverse}
We refer to an edge $(a,b)$ as a "reverse edge" if $a > b$. It is straightforward to calculate the probability that a reverse edge is added at timestep $t$ as $\frac{2E_t}{n(n-1)}$, where $E_t$ is the number of non-reverse edges in the network at time $t$. Moreover, $E_t \approx t$, since it is rare to add a reverse edge to the network. This allows us to estimate the expected total number of reverse edges in the network at time $t$, $R_t$, as follows:

\begin{eqnarray*}
R_t &\approx& \sum \frac{2i}{n(n-1)}  \\ 
&=&\frac{2}{n(n-1)} \sum i = \frac{2}{n(n-1)}\frac{t(t+1)}{2} = \frac{t(t+1)}{n(n+1)}
\end{eqnarray*}

Hence if $t=dn$, then $R_t \approx \frac{t^2}{n^2} = \delta^2$ . That is, the expected number of reverse edges increases as the square of the edge density.

\subsection{Estimating the out-component distribution without reverse edges}
\label{out}

In this section we introduce a branching process which approximates the out-component of a single starting node with a given rank. The branching process does not factor in reverse edges, but we will show in Sec. VI. C that while the process remains in a subcritical phase, where no giant out-components exist, the reverse edges have an insignificant effect on the component structure. With this in mind, we will delay exploration of the effects of reverse edges, and in this section we will analyze the out-component distribution if the existence of reverse edges is ignored. That is, every directed edge of the the form $(j,i)$ where $j > i$ are removed, leaving only the directed edge $(i,j)$. This will also give insight into the in-component distribution, which mirrors the out-component distribution. Since by reversing the directionality of the edges the out-components become in-components and vice versa, it is straightforward to determine that the out-component distribution of the node ranked $i$ is the same as the in-component distribution of the node ranked $n - i + 1$. In symbols, $\Pr(OUT(i) = k) = \Pr(IN(n - i + 1) = k)$.  

We define a multitype branching process in which each node is labeled and each parent node labeled $i$ gives birth to a single child node labeled $j$ with fixed probability $p$ for all $j>i$. Additionally, there is a maximum label $n$. One somewhat unusual property of this branching process is that it is guaranteed to terminate regardless of the value chosen for $p$, since a parent always gives birth to children of strictly higher label number, resulting in a maximum depth of $n$. With a carefully chosen value of $p$ which depends on the edge density of the ODER process, this branching process approximates the out-component distribution of a single node of the ODER process, replacing the labels with ranks. To avoid confusion, we use the word ``label" when referring to a node in the branching process and the word "rank" when referring to a node in the ODER process. Note that in the ODER process, the probability that the edge $(i,j)$ exists for $i < j$ after 1 edge addition is $\frac{2}{n(n-1)}$. Hence the probability that the edge $(i,j)$ exists after $m$ edge additions is $1 - (1 - \frac{2}{n(n-1)})^m$. So to use the branching process defined in this section to approximate the out-component distribution of a single node after $m$ edge additions, we take $p =1 - (1 - \frac{2}{n(n-1)})^m$.  

For any branching process, we define the "total progeny" of a node labeled $i$, $T_i$, as the number of nodes descended from that node together with the node itself. The total progeny corresponds to the size of the out-component distribution of the node ranked $i$ in the ODER process. If the initial node of the branching process is $i$, the total progeny approximates the number of nodes contained in the out-component of $i$. For convenience, we define a collection of variables $T_j^i$, with $T_j^i$ identically distributed to $T_j$, which in the equation below denotes the total progeny of a child with label $j$ whose parent is labeled $i$. Finally, we define $I_j$ as a set of i.i.d. identity variables which output $1$ with probability $p$ and $0$ otherwise. This yields the following relation for $T_i$:

\begin{eqnarray*}
T_i &=& 1 + \sum_{j = i + 1}^n I_j T_j^i. 
\end{eqnarray*}

Note that $T_n = 1$, since a parent with label $n$ can have no children due to the way the process has been defined. Using this relation, we may solve for the expected value of the total progeny of this branching process starting with a parent of any given label. 

\begin{eqnarray}
\langle T_i \rangle &=& 1 + \sum_{j = i + 1}^n \langle I_j \rangle \langle T_j \rangle = 1 + \sum_{j = i + 1}^n p\langle T_j \rangle \\
\langle T_n \rangle &=& T_n = 1
\end{eqnarray}

Note that $\langle T_i \rangle$ is a real-valued function which may be defined as the unique solution (1). This solution can be obtained exactly or closely approximated by replacing the sums with integrals and solving the resultant integral equations using the fundamental theorem of calculus to turn them into first order linear differential equations. In general, it is straightforward to solve an equation of the form $y = \alpha + \int_x^n \gamma y \mathrm{d} y$ for any fixed constants $\alpha$ and $\gamma$ using the fundamental theorem of calculus as follows:

\begin{eqnarray*}
y &=& \alpha + \int_x^n \gamma y\mathrm{d} y = \alpha + \gamma[Y(n) - Y(x)] \\
y' &=& - \gamma y(x) = - \gamma y \\
\frac{y'}{y} &=& - \gamma \\
\log y &=& - \gamma x + C \\
y &=& De^{- \gamma x}
\end{eqnarray*}

Using the above formula along with the condition $\langle T_n \rangle=1$ , we find that $\langle T_i \rangle =  e^{pn}e^{- pi} = e^{p(n - i)}$. Moreover, it is easily checked that the exact solution is given by $\langle T_{n - i} \rangle = (1 + p)^i$ or equivalently $\langle T_{i} \rangle = (1 + p)^{n - i}$.

As stated previous, in order to use this branching process to approximate the ODER process, we define $p$ as the independent probability that any given edge (i,j) exists, where $1 \leq i < j \leq n$ after $m$ edges have been independently added uniformly at random. We can then approximate:
\begin{eqnarray*}
p_m &=&  1 - \left( 1 - \frac{2}{n(n-1)}\right)^{dn} \\
&=& 1 - \left(1 - \frac{2}{n(n-1)}\right)^{n(n-1)\frac{d}{n-1}} \\
&\approx& 1 - e^{-\frac{2d}{n-1}} \approx 1 - \left( 1 - \frac{2d}{n} \right) = \frac{2d}{n}
\end{eqnarray*}

The out-component distribution of a node labeled $i$ (ignoring reverse edges) after $m$ edge additions is approximated by the branching process described above with $p$ taken to be $p_m$. In particular, the branching process estimated the expected size of the out-component of node $i$ as $\langle T_i \rangle = (1 + p_m)^{n - i} \approx (1 + \frac{2d}{n})^{n - i} \approx e^{\frac{2d(n - i)}{n}}$. This estimate is a strict upper bound on the expected size of the out-component, since the total progeny of the branching process may include multiple nodes with the same label. It follows that while the density $d$ is constant with respect to the number of nodes n, the expected size of the out-component starting from any node is bounded by a constant, and it is necessary for the density to be of order $\log n$ before the expected size of any out-component can be of order $n$.

\subsection{Effects of reverse edges on component structure}
\label{effects}

In this section, we will show that reverse edges gave an insignificant effect when no giant out-components exist, forming only a small number of tiny SCCs. First, suppose that the density is low enough that with high probability no giant out-components exist in the network. This implies that there are $O(\log^2(n))$ reverse edges. Moreover, since the largest out-component is with high probability $O(n^\gamma)$ for some $0< \gamma < 1$, the same can be said of the largest SCC. It follows that with high probability each SCC is the result of the addition of a single reverse edge. That is, there is no SCC which contains multiple reverse edges. 

Suppose the edge $(a,b)$ exists and the reverse edge $(b,a)$ is about to be added, and that there are no other reverse edges in the current strongly connected components containing $a$ or $b$. In this case, the resultant strongly connected component will contain all nodes in the strongly connected components of $a$ and $b$ along with those in the intersection of the out-component of $a$ and the in-component of $b$, which are necessarily in the interval $[a,b]$. The out-component of node $a$ is skewed towards higher ranked nodes while the in-component of node b is skewed towards lower ranked nodes, so the resultant SCC is actually smaller in expected value than $OUT_{[a,b]}(a)IN_{[a,b]}(b)$, possibly much smaller.

If the largest out-component or in-component is $O(n^\gamma)$ for some $0< \gamma < 1$, then with high probability the out-components and in-components of both nodes linked by each reverse edge are distinct, and hence each reverse edge can be added independently without need to consider the effect of previous reverse edge added. After adding on $\log(n)^2$ reverse edges, the expected number of nodes that are in any in-component or out-component of any head or tail node of any reverse edge will be $O(\log(n)^2 n^\gamma) = o(n)$. That is, almost all nodes are isolated with respect to the SCC structure, having in-components and out-components whose intersection is only a single node. 

This indicates that in the thermodynamic limit giant in-components and out-components must emerge before non-trivial strong component structure exists on a non-vanishing set of nodes. This is a necessary condition for the discontinuous emergence of a giant strongly connected component, which must occur via the addition of a reverse edge $(a,b)$ in which the intersection of the out-component of $b$ with the in-component of $a$ is giant.

\section{Conclusion}

In this paper we have analyzed the percolation behavior of the ODER and C-ODER processes on ordered, directed graphs. We gave numerical and analytical evidence showing that the largest jump in the size of the largest strongly connected component in the ODER process is proportional to system size, indicating a discontinuous jump in the size of the largest strongly connected component resulting from the merging of two giant strongly connected components. Fig. 7 demonstrates that the addition of a single edge is enough to result in a discontinuous jump in the size of the largest strongly connected component, and that the relative size of this jump does not decrease as the system size increases. Fig. \ref{fig:fig7million} shows that this is the result of merging two existing giants.

The addition of a competitive mechanism in the C-ODER process amplifies the size of the discontinuous nature of the jump, from around $5\%$ of system size to around $\frac{1}{3}$ of system size. This is a significant advance from previous work which only exhibited discontinuous jumps in sizes of the largest in-component, out-component, and weakly connected component.\cite{Squires13} Moreover, we have shown that the basic mechanism used has a basis in reality, with $99\%$ percent of the edges in the Higgs Boson retweet network being reverse edges. This may lead to a far greater understanding of how to control and/or predict explosive percolation on real-world directed networks.  However, in order to be applicable, there must exist a natural ordering with respect to which activity tends to flow monotonically. It is intuitive to believe that on most directed online social networks  this would be the case if users are ranked, for instance, by the number of times their content is shared. Similar phenomena may occur in other types of networks as well, such as function call networks in which functions are ranked by how often they are called by other functions~\cite{HaoranWen_apache}. A useful metric is the number of reverse edges, which varies depending on the ordering used. Hence if it is unclear how to define an ordering, one approach could be to find the ordering which minimizes the number of reverse edges. Conversely, if there is an obvious ordering to use, it could be shown to nearly minimize the number of reverse edges among all possible orderings. 

We found that the overlap(as defined in Sec. \ref{interval}) between the multiple co-existing giant SCCs arising from the C-ODER process is very small. This indicates that the two giant SCCs separate the nodes by ranking, with one component of low rank users and one component of high rank users. This novel behavior is not well understood, and it is is desirable both to understand why this separation in ranking occurs and to observe it in real world networks.  

We used branching process analysis to explain the high edge density required for a giant SCC to emerge, and to give basic intuition for the percolation processes. The branching processes used here have vastly different behavior from those usually used to analyze random networks. They are multitype branching processes in which the number of types is equal to the number of nodes, and with the children always having fewer children on average than their parents. This leads to certain extinction regardless of the parameters used. Therefore, instead of studying the survival probability, we study the total progeny in order to place bounds on the size of the largest out-component and in-component. A limitation of these models is the assumption of monotonicity; all edges added in the subcritical phase are reverse edges. Similar branching processes could be used to study percolation processes which drop the assumption of monotonicity, in which there is a small probability that each edge added is not a reverse edge.   

{\bf Acknowledgements:} We thank Pierre-Andr\'e No\"el for useful discussions. We gratefully acknowledge support from the U.S. Army Research Office under Multidisciplinary University Research Initiative Award No. W911NF-13-1-0340, and Cooperative Agreement No. W911NF-09-2-0053, and  the U. S. Department of Defense, Minerva grant No. W911NF-15-1-0502.

\bibliographystyle{unsrt}
\bibliography{dirEP}

\end{document}